\documentclass[%
 reprint,
 amsmath,amssymb,
 aps,
 nofootinbib,
 superscriptaddress,
 longbibliography
]{revtex4-2}

\usepackage{bm}
\usepackage{braket}
\usepackage{amsmath}
\usepackage{amssymb}
\usepackage{amsthm}
\usepackage{enumerate}
\usepackage{booktabs}
\usepackage{txfonts}
\usepackage{algcompatible}
\usepackage{algorithm}
\usepackage{tikz}
\usetikzlibrary{quantikz}
\usepackage{subfigure}
\usepackage{caption}
\usepackage[
colorlinks=true,
urlcolor=blue,
citecolor=blue,
linkcolor=blue,
hyperfootnotes=false]{hyperref}

\newcommand\EatDot[1]{}

\makeatletter
\def\hlinewd#1{%
	\noalign{\ifnum0=`}\fi\hrule \@height #1 %
	\futurelet\reserved@a\@xhline}
\makeatother

\begin{document}

\renewcommand{\thefootnote}{}

\preprint{APS/123-QED}

\title{Extracting a function encoded in amplitudes of a quantum state by tensor network and orthogonal function expansion}

\author{Koichi Miyamoto}
\email{miyamoto.kouichi.qiqb@osaka-u.ac.jp}
\affiliation{Center for Quantum Information and Quantum Biology, Osaka University, Toyonaka, Osaka 560-0043, Japan}

\author{Hiroshi Ueda}
\email{ueda.hiroshi.qiqb@osaka-u.ac.jp}
\affiliation{Center for Quantum Information and Quantum Biology, Osaka University, Toyonaka, Osaka 560-0043, Japan}
\affiliation{JST, PRESTO, Kawaguchi, Saitama 332-0012, Japan}
\affiliation{Computational Materials Science Research Team, RIKEN Center for Computational Science (R-CCS), Kobe 650-0047, Japan}

\date{\today}

\begin{abstract}
There are quantum algorithms for finding a function $f$ satisfying a set of conditions, such as solving partial differential equations, and these achieve exponential quantum speedup compared to existing classical methods, especially when the number $d$ of the variables of $f$ is large.
In general, however, these algorithms output the quantum state which encodes $f$ in the amplitudes, and reading out the values of $f$ as classical data from such a state can be so time-consuming that the quantum speedup is ruined.
In this study, we propose a general method for this function readout task.
Based on the function approximation by a combination of tensor network and orthogonal function expansion, we present a quantum circuit and its optimization procedure to obtain an approximating function of $f$ that has a polynomial number of degrees of freedom with respect to $d$ and is efficiently evaluable on a classical computer.
We also conducted a numerical experiment to approximate a finance-motivated function to demonstrate that our method works.
\end{abstract}

\footnote{Corresponding author: Koichi Miyamoto}

\pacs{Valid PACS appear here}
                              
\maketitle

\renewcommand{\thefootnote}{\arabic{footnote}}
\setcounter{footnote}{0}

\section{\label{sec:intro}Introduction}

Quantum computing is an emerging technology that is expected to provide speedup for various classically time-consuming problems.
Following recent
advances, researchers are now investigating its practical applications.
Some fundamental quantum algorithms have evolved into solvers for more concrete numerical problems.
For example, the Harrow-Hassidim-Lloyd algorithm \cite{harrow2009quantum} to solve linear equation systems and its extensions \cite{ambainis2012variable,clader2013preconditioned,childs2017quantum,chakraborty2019power,subacsi2019quantum,lin2020optimal,tong2021fast,an2022quantum,costa2021optimal} led to quantum solvers for ordinary differential equations (ODEs) \cite{berry2014high,berry2017quantum,childs2020quantum,xin2020quantum} and partial differential equations (PDEs) \cite{cao2013quantum,clader2013preconditioned,montanaro2016quantum,costa2019quantum,childs2021high,linden2020quantum}, including even nonlinear systems \cite{liu2021efficient,xue2021quantum,lloyd2020quantum,krovi2022improved,jin2022quantum,an2022efficient}.
The practical use of these algorithms has been proposed in various fields such as epidemiology \cite{kiani2022quantum,liu2021efficient}, fluid dynamics \cite{liu2021efficient}, and financial derivative pricing \cite{miyamoto2021pricing}.
Compared with existing classical methods, these quantum algorithms achieve exponential speedup with respect to the size of the problem.
For an ODE, the problem size is the number of equations in the system; for a PDE, it is the dimension $d$ of the domain of the solution function $f$, that is, the number of the variables of $f$.
Thus, they are among the most promising applications of quantum computing.
In addition to the above algorithms that run on fault-tolerant quantum computers (FTQCs), some algorithms for noisy intermediate-scale quantum (NISQ) devices also solve ODEs and PDEs \cite{endo2020variational,lubasch2020variational,liu2021variational,sato2021variational,fontanela2021quantum,kyriienko2021solving,Leong2022VariationalQE,garcia2021solving,alghassi2022variational,gonzalez2021pricing,radha2021quantum,joo2021quantum,paine2022quantum,kubo2022pricing,liu2022application}.

However, when we consider these and other quantum solvers and the speedup they provide, we should pay attention to the meaning of the words ``solve" and ``speedup".
That is, we must resolve the issue of {\it how to extract the function values from the quantum state}, which might ruin the quantum speedup.

The detail is as follows.
As pointed out in \cite{aaronson2015read}, many quantum algorithms output the solution not as a figure that we eventually want, but as the quantum state in which the solution is encoded as the amplitudes.
For example, quantum algorithms for solving a PDE output the quantum state in the form of 
\begin{equation}
\ket{f}:=\frac{1}{C}\sum_{i=0}^{N_{\rm gr}-1} f(\vec{x}_i)\ket{i}~,
\end{equation}
where $f$ is the solution function, $\vec{x}_0,\vec{x}_1,...,\vec{x}_{N_{\rm gr}-1}$ are the $N_{\rm gr}$ grid points in the domain of $f$, $\ket{0},\ket{1},...,\ket{N_{\rm gr}-1}$ are the computational basis states, and $C$ is the normalization factor.
Reading out the values of the function as classical data from such a quantum state can often be a bottleneck.
For the example of solving a PDE, the number
of grid points $N_{\rm gr}$ is exponentially large with respect to the dimension $d$: naively, taking $n_{\rm gr}$ grid points in each dimension leads to $N_{\rm gr}=n_{\rm gr}^d$ points in total. 
This makes the amplitude $\frac{1}{C}f(\vec{x}_i)$ of each basis state in $\ket{f}$ exponentially small when the amplitude is not localized on certain grids.
This means that when we try to retrieve the amplitude and then the function value $f(\vec{x}_i)$ at the point $x_i$ using methods such as quantum amplitude estimation \cite{brassard2002,suzuki2020amplitude}, an exponentially large time overhead is added.
Therefore, even if a quantum algorithm exists that generates the state $\ket{f}$ exponentially faster than a classical algorithm outputs the value of $f(\vec{x}_i)$, obtaining $f(\vec{x}_i)$ through the quantum algorithm might not be faster than the classical one.

If we want to obtain the function values at many points, for example, to plot the values as a graph, the situation worsens.
To obtain the function values at $M$ points, the quantum algorithm must be repeated $M$ times.

Motivated by this background, this study focuses on how to efficiently extract the function encoded as the amplitudes in the quantum state $\ket{f}$, given an oracle $O_f$ to generate it.
Although we can resolve this issue using the specific nature of the problem in some cases such as derivative pricing considered in \cite{miyamoto2021pricing}, where the martingale property of the derivative price is used to calculate the function value at a point, nevertheless, it is desirable to devise methods that are generally applicable.

The method proposed in this study is twofold.
First, we use an {\it orthogonal function} system.
Orthogonal functions such as trigonometric functions, Legendre polynomials, Chebyshev polynomials, and so on are the sequences of functions orthogonal to each other with respect to some inner product, which is defined as the weighted integral of the product of two functions over the domain or the sum of the values of the product at the specific nodes.
Orthogonal functions are often used for function approximation.
Any function $f$ satisfying some conditions of smoothness is approximated as $f\approx \sum_l a_l P_l$, the series of orthogonal functions $\{P_l\}_l$, with the coefficient $a_l$ given by the inner product of $f$ and $P_l$. 
We expect that orthogonal function expansion may also be used in the quantum setting, because, as explained later, the coefficient $a_{l}$ is given by $\braket{P_l|f}$ multiplied by a known factor, with $\ket{P_l}$ being the quantum state in which $P_l$ is encoded like $\ket{f}$.
Thus, by estimating $\braket{P_l|f}$ for every $l$ up to a sufficiently high order, we obtain the orthogonal function expansion $\tilde{f}$ of $f$, and then the approximate values of $f$ at arbitrary points by evaluating $\tilde{f}$. 
This approach seems promising 
because we expect that $\braket{P_l|f}$ is not exponentially small, 
unlike the amplitudes of the computational basis states in $\ket{f}$ (see Sec. \ref{sec:oracle}).

However, in the high-dimensional case, the above approach still suffers from high complexity.
If we use the $D$ orthogonal functions $\{P_l\}_{l\in[D]_0}$\footnote{For $n\in\mathbb{N}$, we define $[n]_0:=\{0,1,...,n-1\}$.} for an accurate approximation in the one-dimensional case, the naive way to achieve similar accuracy in the $d$-dimensional case is 
to use the tensorized functions $\{P_{\vec{l}}\}_{\vec{l}\in[D]_0^d}$, where $P_{\vec{l}}(x_1,...,x_d)=\prod_{i=1}^dP_{l_i}(x_i)$ for $\vec{l}=(l_1,...,l_d)$.
In this way, because the total number of $P_{\vec{l}}$'s is $D^d$, obtaining the coefficients $a_{\vec{l}}$ for all $P_{\vec{l}}$'s and then the orthogonal function expansion of $f$ exhibits exponential complexity, and so does evaluating the resultant expansion.
Although this is less serious than reading out the amplitudes in $\ket{f}$ because we take $D<n_{\rm gr}$, the exponential dependence of the complexity on $d$ still exists.

Then, we make use of the second building block of our method: {\it tensor network}, especially the type called {\it matrix product state (MPS)}, which is simple and widely used (for reviews, see \cite{orus2014practical, okunishi2022developments}).
Tensor network is an approximation scheme for high-order tensors as a contraction of lower-order tensors.
In some situations, it approximates the original tensor well, reducing the degrees of freedom (DOF) and data volume.
It was originally invented in quantum many-body physics to approximate wave functions in intractably high-dimensional Hilbert spaces; however, it is currently used to reduce complexity in various fields including function approximation \cite{nouy2017low,griebel2021analysis,ali2020approximation1,ali2020approximation2,ali2021approximation,bachmayr2021approximation,griebel2022low}.
A recent study \cite{griebel2022low} showed that the complexity of the tensor network approximation of a $d$-dimensional function does not scale exponentially on $d$ under certain conditions, which indicates the powerful approximation ability of tensor network.

Another advantage of tensor network is its compatibility with quantum computing.
That is, we can generate a quantum state in which a type of tensor network is amplitude-encoded using a simple quantum circuit \cite{ran2020encoding}.
Moreover, there is a general procedure for optimizing such a tensor network circuit to maximize the fidelity between the resulting state and a given state \cite{shirakawa2021}.
Therefore, we reach the following idea: given $O_f$, we find a tensor network approximation of the coefficients $a_{\vec{l}}$ in the orthogonal function expansion of $f$ and then an approximate function of $f$ through quantum circuit optimization.

In the remainder of this paper, we show how this idea is concretely realized.
We present the tensor-network-based quantum circuit and optimization procedure for making the generated state close to $\ket{f}$, based on \cite{shirakawa2021}.
The parameters in the tensor network are easily read out from the circuit.
Note that this method can be used on both fault-tolerant quantum computers and noisy intermediate-scale quantum computers.

The remainder of this paper is organized as follows.
Sec. \ref{sec:method} presents our method.
Beginning with an explanation of the problem setting under consideration, we present the quantum circuit we use and the procedure to optimize it.
Sec. \ref{sec:NumExp}, describes the numerical experiment we conducted to validate our method, which demonstrates that the method successfully approximates a finance-motivated multivariate function.
Sec. \ref{sec:sum} summarizes this paper.

\section{Our method \label{sec:method}}

\subsection{Overview}

\begin{figure}[tp]
\begin{center}
\includegraphics[scale=0.85]{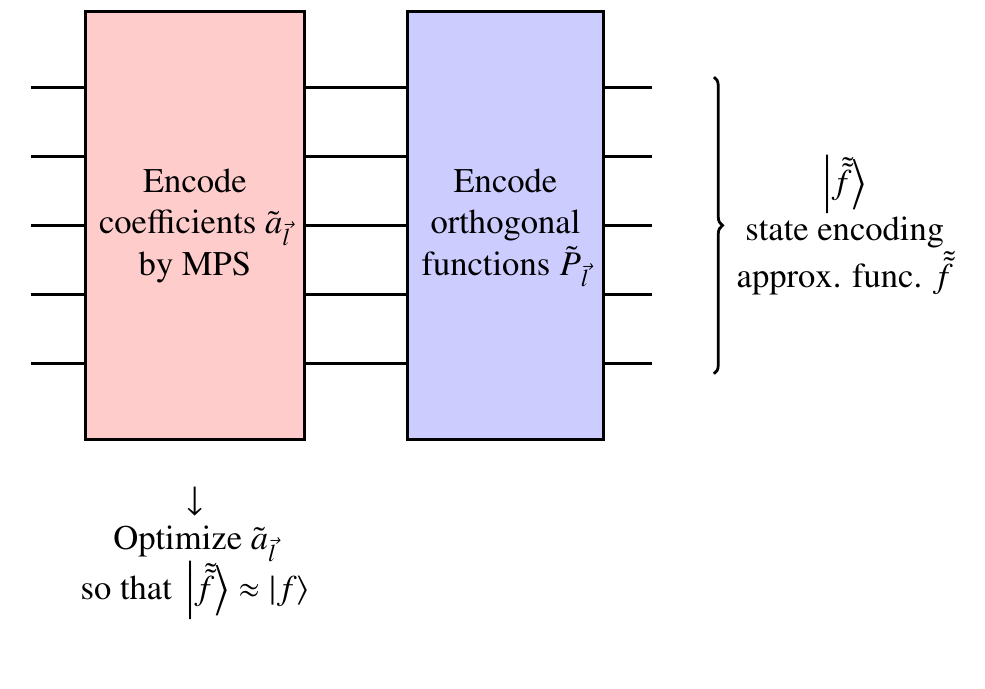}
\caption{Overview of the quantum circuit used in the proposed method.}
\label{fig:overview}
\end{center}
\end{figure}

Before providing a detailed explanation of the proposed method, we present a high-level overview.
For the given function $f$, our goal is to obtain an approximation function efficiently computable on a classical computer.
For this purpose, we use a quantum circuit, whose overview is shown is given in Fig. \ref{fig:overview}.
This consists of two parts.
In the first half, we encode the coefficients $\tilde{a}_{\vec{l}}$ of the orthogonal function series written in the form of an MPS into the amplitudes of the quantum state.
In the latter half, we operate the quantum circuit that encodes the values of the orthogonal functions into the amplitudes.
These two operations yield the quantum state $\Ket{\tilde{\tilde{f}}}$, in which the approximation function $\tilde{\tilde{f}}$ based on the MPS and orthogonal function expansion is encoded in the amplitudes. 
Then, we optimize the first half of the circuit such that $\Ket{\tilde{\tilde{f}}}$ approaches $\Ket{f}$, the state encoding $f$.
As a result, we obtain the coefficients in the form of an MPS optimized to approximate $f$, and then the approximation function $\tilde{\tilde{f}}$, which is efficiently computable by virtue of the MPS.

The approximation problem addressed in this study is explained in more detail in Sec. \ref{sec:problem} and the setting of the oracles available is shown in Sec. \ref{sec:oracle}.
The MPS approximation of the coefficients is explained in Sec. \ref{sec:MPS}, and encoding it into the quantum state, that is, the first half of the circuit, is described in Sec. \ref{sec:QC}.
The entire circuit including the latter half and optimization of the MPS are explained in Sec. \ref{sec:Opt}.

\subsection{Problem description\label{sec:problem}}

As the approximation target, we consider a real-valued function $f$ on the $d$-dimensional hyperrectangle $\Omega:=[L_1,U_1]\times\cdots\times[L_d,U_d]$, where, for $i\in[d]$~\footnote{For $n\in\mathbb{N}$, we define $[n]:=\{1,...,n\}$.}, $L_i$ and $U_i$ are real values such that $L_i<U_i$.

For any $i\in[d]$, we also consider orthogonal functions $\{P^i_l\}_l$ on $[L_i,U_i]$ labeled by $l\in\mathbb{N}_0:=\{0\}\cup\mathbb{N}$.
These functions are characterized by the orthogonal relation that, for any $l,l^\prime\in\mathbb{N}_0$,
\begin{equation}
    \int_{L_i}^{U_i} P^i_{l}(x)P^i_{l^\prime}(x)w^i(x)dx=\delta_{l,l^\prime}, \label{eq:orthoInt}
\end{equation}
where the weight function $w^i$ is defined on $[L_i,U_i]$ and takes a non-negative value, and $\delta_{l,l^\prime}$ is the Kronecker delta.
However, we hereafter assume that $\{ P^i_l \}_l$ satisfy the discrete orthogonal relation as follows: for any $D\in\mathbb{N}$, there exist $n_{\rm gr}$ points $x_{i,0}<x_{i,1}<...<x_{i,n_{\rm gr}-1}$ in $[L_i,U_i]$, 
where $n_{\rm gr}\ge D$, such that, for any $l,l^\prime\in[D]_0$,
\begin{equation}
    \sum_{j=0}^{n_{\rm gr}-1} P^i_{l}(x_{i,j})P^i_{l^\prime}(x_{i,j})=c^i_l\delta_{l,l^\prime}
    \label{eq:orthoDisc}
\end{equation}
holds for some $c^i_l>0$.
We can consider Eq. (\ref{eq:orthoDisc}) as the discrete approximation of Eq. (\ref{eq:orthoInt}), with $w^i$ absorbed into the spacing of the grid points $x_{i,j}$.
For some orthogonal functions such as trigonometric functions and Chebyshev polynomials, the relationship in Eq. (\ref{eq:orthoDisc}) holds strictly.

We define the tensorized orthogonal functions as follows: for any $\vec{l}=(l_1,...,l_d)\in \mathbb{N}_0^d$ and $\vec{x}=(x_1,...,x_d)\in\mathbb{R}^d$,
\begin{equation}
    P_{\vec{l}}(\vec{x}):=\prod_{i=1}^d P^i_{l_i}(x_i).
\end{equation}
It follows from Eq. (\ref{eq:orthoDisc}) that, with $\vec{x}_{\vec{j}}$ defined as $\vec{x}_{\vec{j}}:=(x_{1,j_1},...,x_{d,j_d})\in\mathbb{R}^d$ for any $\vec{j}=(j_1,...,j_d)\in[n_{\rm gr}]_0^d$, they satisfy the orthogonal relation
\begin{equation}
    \sum_{\vec{j}\in[n_{\rm gr}]_0^d}P_{\vec{l}}\left(\vec{x}_{\vec{j}}\right)P_{\vec{l}^\prime}\left(\vec{x}_{\vec{j}}\right)=c_{\vec{l}}\delta_{\vec{l},\vec{l}^\prime} \label{eq:ortho}
\end{equation}
for any $\vec{l}=(l_1,...,l_d)$ and $\vec{l}^\prime=(l_1^\prime,...,l_d^\prime)$ in $[D]_0^d$, where $\delta_{\vec{l},\vec{l}^\prime}:=\prod_{i=1}^d \delta_{l_i,l_i^\prime}$ and $c_{\vec{l}}:=\prod_{i=1}^d c_{l_i}^i$.

We can use these ${P_{\vec{l}}}_{\vec{l}}$ to approximate $f$ by the orthogonal function series
\begin{equation}
    \tilde{f}(\vec{x}) := \sum_{\vec{l}\in[D]_0^d} a_{\vec{l}}P_{\vec{l}}(\vec{x}) \label{eq:orthoExp}
\end{equation}
with $D$ such that $\max_{\vec{x}} |f(x)-\tilde{f}(x)| $ is less than the desired threshold $\epsilon$.
Here, for any $\vec{l}\in[D]_0^d$, the coefficient $a_{\vec{l}}$ is given by
\begin{equation}
    a_{\vec{l}} =  \frac{1}{c_{\vec{l}}}\sum_{\vec{j}\in[n_{\rm gr}]_0^d} f\left(\vec{x}_{\vec{j}}\right)P_{\vec{l}}\left(\vec{x}_{\vec{j}}\right), \label{eq:coef}
\end{equation}
which follows from Eq. (\ref{eq:orthoDisc}).
It is known that this kind of series converges to $f$ in the large $D$ limit for some orthogonal functions (see \cite{zhizhiashvili1996trigonometric} for the trigonometric series and \cite{trefethen2019approximation} for the Chebyshev series).

\subsection{Oracles to generate the function-encoding states \label{sec:oracle}}

In this study, we assume the availability of the oracle $O_f$ mentioned in the introduction.
We now define this formally.
Hereafter, we assume that the grid number $n_{\rm gr}$ satisfies $n_{\rm gr}=2^{m_{\rm gr}}$ with some integer $m_{\rm gr}$.
Then, we consider the system $S$ consisting of $d m_{\rm gr}$-qubit registers and the unitary operator $O_f$ that acts on $S$ as
\begin{equation}
    O_f \ket{0}^{\otimes d}=\ket{f}:=\frac{1}{C} \sum_{\vec{j}=(j_1,...,j_d)\in[n_{\rm gr}]_0^d} f\left(\vec{x}_{\vec{j}}\right) \ket{j_1}\cdots\ket{j_d}. \label{eq:Of}
\end{equation}
For any $n\in\mathbb{N}_0$, we denote by $\ket{n}$ the computational basis states on the quantum register with a sufficient number of qubits, in which the bit string on the register corresponds to the binary representation of $n$. 
$\ket{j}$ with $j\in[n_{\rm gr}]_0$ in Eq. (\ref{eq:Of}) is the computational basis state on an $m_{\rm gr}$-qubit register.
Furthermore, 
$C$ in Eq. (\ref{eq:Of}) is defined as
\begin{equation}
    C:= \sqrt{\sum_{\vec{j}\in[n_{\rm gr}]_0^d} \left(f\left(\vec{x}_{\vec{j}}\right)\right)^2}. \label{eq:C}
\end{equation}

We also assume that the availability of the oracles $V^1_{\rm OF},...,V^d_{\rm OF}$, each of which acts on an $m_{\rm gr}$-qubit register as
\begin{equation}
    V^i_{\rm OF}\ket{l} = \ket{P^i_l} := \frac{1}{\sqrt{c^i_l}}\sum_{j=0}^{n_{\rm gr}-1} P^i_{l}(x_{i,j})\ket{j} \label{eq:ViOF}
\end{equation}
for any $l\in[D]_0$.
$V^i_{\rm OF}\ket{D},...,V^i_{\rm OF}\ket{n_{\rm gr}-1}$ may be any states as far as $V^i_{\rm OF}$ is unitary.
In principle, these oracles are constructible because of the orthogonal relation (\ref{eq:orthoDisc}).
We discuss their implementation in Appendix \ref{sec:ViOF}.

Now, let us elaborate on the reason why reading out $f\left(\vec{x}_{\vec{j}}\right)$ for $\vec{j}\in[n_{\rm gr}]_0^d$ from $\ket{f}$ is difficult but obtaining it through orthogonal function expansion is more promising, as mentioned briefly in the introduction.
We rewrite the amplitude in $\ket{f}$ as
\begin{equation}
    \frac{f\left(\vec{x}_{\vec{j}}\right)}{C}=\frac{1}{\sqrt{N_{\rm gr}}}\frac{f\left(\vec{x}_{\vec{j}}\right)}{\bar{f}},
\end{equation}
where $\bar{f}:=\sqrt{\frac{1}{N_{\rm gr}}\sum_{\vec{j}\in[n_{\rm gr}]_0^d} \left(f\left(\vec{x}_{\vec{j}}\right)\right)^2}$ is the root mean square of $f$ over the grid points.
The amplitude is suppressed by the factor $1/\sqrt{N_{\rm gr}}$, which is exponential with respect to $d$, and thus exponentially small unless $f$ is extremely localized at $\vec{x}_{\vec{j}}$ such that $f\left(\vec{x}_{\vec{j}}\right)/\bar{f}$ is comparable to $\sqrt{N_{\rm gr}}$.
However, we note that, for any $\vec{l}\in[D]_0^d$, we have
\begin{equation}
    \Braket{P^i_{\vec{l}} | f} = \frac{1}{C\sqrt{c_{\vec{l}}}}\sum_{\vec{j}\in[n_{\rm gr}]_0^d} f\left(\vec{x}_{\vec{j}}\right)P^i_{\vec{l}}\left(\vec{x}_{\vec{j}}\right) = \frac{\sqrt{c_{\vec{l}}}}{C}a_{\vec{l}},
\end{equation}
where $\Ket{P^i_{\vec{l}}}:=\ket{P^1_{l_1}}\cdots\ket{P^d_{l_d}}$.
Because both $C$ and $\sqrt{c_{\vec{l}}}$ are the root mean squares of some functions over the grid points, we expect that their ratio is $O(1)$ and that we can efficiently obtain the expansion coefficient $a_{\vec{l}}$ and then the approximation $\tilde{f}$ of $f$ in Eq. (\ref{eq:orthoExp}) by estimating $\Braket{P^i_{\vec{l}} | f}$, without suffering from an exponential suppression factor such as $1/\sqrt{N_{\rm gr}}$.
However, we do not consider this direction in this study, because finding all the expansion coefficients suffers from an exponential increase in the number of coefficients in the high-dimensional case, as explained in the introduction.

\subsection{Matrix product state \label{sec:MPS}}

Thus, we consider using a tensor network.
Among the various types of tensor networks, we use MPS, also known as the {\it tensor train}, which is simple but powerful and therefore widely used in various fields; a basic introduction of the MPS is presented in Appendix~\ref{sec:basic_mps}. 
In MPS scheme, the order-$d$ tensor $a_{\vec{l}}\in\mathbb{R}^{\overbrace{D\times\cdots\times D}^d}$ is approximated by
\begin{equation}
    \tilde{a}_{\vec{l}} := \sum_{k_1=1}^{r}\cdots\sum_{k_{d-2}=1}^{r} U^{1}_{l_1,k_1} U^{2}_{k_1,l_2,k_2} \cdots U^{d-2}_{k_{d-3},l_{d-2},k_{d-2}} U^{d-1}_{k_{d-2},l_{d-1},l_d}, \label{eq:MPS}
\end{equation}
where $r\in\mathbb{N}$ is called the bond dimension, $U^1\in\mathbb{R}^{D \times r}$, $U^i\in\mathbb{R}^{r\times D \times r}$ for $i\in\{2,...,d-2\}$, and $U^{d-1}\in\mathbb{R}^{r \times D\times D}$ \footnote{Usually, the MPS representation of a $d$-dimensional tensor takes the form
\begin{equation}
    \tilde{a}_{\vec{l}} := \sum_{k_1=1}^{r}\cdots\sum_{k_{d-1}=1}^{r} U^{1}_{l_1,k_1} U^{2}_{k_1,l_2,k_2} \cdots U^{d-1}_{k_{d-2},l_{d-1},k_{d-1}} U^{d}_{k_{d-1},l_{d}}, \label{eq:MPSUsual}
\end{equation}
where, in comparison to Eq. (\ref{eq:MPS}), $(U^{d-1}_{k_{d-2},l_{d-1},k_{d-1}})$ is in $\mathbb{R}^{r \times D\times r}$ rather than $\mathbb{R}^{r \times D\times D}$, and $(U^d_{k_{d-1}l_d})\in\mathbb{R}^{r \times D}$ is added.
We can consider that $U^{d-1}$ and $U^d$ in Eq. (\ref{eq:MPSUsual}) are contracted to $U^{d-1}$ in Eq. (\ref{eq:MPS}).
The reason for the form in Eq. (\ref{eq:MPS}) is that it corresponds to the quantum circuit considered in Sec. \ref{sec:QC}.
In addition, although we can set the bond dimension $r$ separately for each pair of $U^i$ and $U^{i+1}$, for simplicity we set it to the same value.
}.
We also impose the following conditions: for any $i\in\{2,...,d-2\}$ and $k_{i-1},k_{i-1}^\prime\in[r]$,
\begin{equation}
    \sum_{l_i=0}^{D-1}\sum_{k_i=1}^r U^i_{k_{i-1},l_i,k_i}U^i_{k_{i-1}^\prime,l_i,k_i}=\delta_{k_{i-1},k_{i-1}^\prime} \label{eq:UiOrtho1}
\end{equation}
holds, and, for any $k_{d-2},k_{d-2}^\prime\in[r]$,
\begin{equation}
    \sum_{l_{d-1}=0}^{D-1}\sum_{l_d=0}^{D-1} U^{d-1}_{k_{d-2},l_{d-1},l_d}U^{d-1}_{k_{d-2}^{\prime},l_{d-1},l_d}=\delta_{k_{d-2},k_{d-2}^\prime} \label{eq:UiOrtho2}
\end{equation}
holds.
We call this form of the MPS the right canonical form.
This can always be imposed by the procedure explained in Appendix~\ref{sec:basic_mps}. 


The representation in Eq. (\ref{eq:MPS}) actually reduces the DOF compared with the original $a_{\vec{l}}$ as a $d$-dimensional tensor.
The total number of components in $U^1,...,U^{d-1}$ is
\begin{eqnarray}
rD + (d-3)r^2D + rD^2,
\end{eqnarray}
which is smaller than that in $a_{\vec{l}}$, $D^d$, unless $r=O(D^{O(d)})$.
Furthermore, with the coefficients $a_{\vec{l}}$ in Eq. (\ref{eq:orthoExp}) represented as Eq. (\ref{eq:MPS}), the computation of the approximation of the function $f$ becomes efficient.
We now approximate $f$ by
\begin{widetext}
\begin{equation}
    \tilde{\tilde{f}}(\vec{x})=\sum_{\vec{l}\in[D]_0^d} \tilde{a}_{\vec{l}} P_{\vec{l}}(\vec{x})
    = \sum_{l_1=0}^{D-1}\cdots\sum_{l_{d}=0}^{D-1}\sum_{k_1=1}^{r}\cdots\sum_{k_{d-2}=1}^{r} U^{1}_{l_1,k_1} U^{2}_{k_1,l_2,k_2} \cdots U^{d-2}_{k_{d-3},l_{d-2},k_{d-2}} U^{d-1}_{k_{d-2},l_{d-1},l_d} P^1_{l_1}(x_1)\cdots P^d_{l_d}(x_d). \label{eq:MPSapp}
\end{equation}
This can be computed as
\begin{equation}
    \tilde{\tilde{f}}(\vec{x})=\sum_{k_1=1}^{r}\cdots\sum_{k_{d-2}=1}^{r} \left(\sum_{l_1=0}^{D-1}U^{1}_{l_1,k_1}P^1_{l_1}(x_1)\right)\times\left(\sum_{l_2=0}^{D-1}U^{2}_{k_1,l_2,k_2}P^2_{l_2}(x_2)\right) \times \cdots \times \left(\sum_{l_{d-2}=0}^{D-1}U^{d-2}_{k_{d-3},l_{d-2},k_{d-2}}P^{d-2}_{l_{d-2}}(x_{d-2})\right)\times \left(\sum_{l_{d-1},l_d=0}^{D-1}U^{d-1}_{k_{d-2},l_{d-1},l_{d}}P^{d-1}_{l_{d-1}}(x_{d-1})P^d_{l_d}(x_d)\right), \label{eq:tiltilf}
\end{equation}
\end{widetext}
that is, we first contract $U^i_{k_{i-1},l_i,k_i}$ and $P^i_{l_i}$ with respect to $l_i$ for each $i\in[d]$, and then take contractions with respect to $k_1,...,k_{d-2}$.
In this procedure, the number of arithmetic operations is
\begin{equation}
    O(dr^2D+rD^2),
\end{equation}
which is much smaller than $O(D^d)$ for computing (\ref{eq:orthoExp}) with general $a_{\vec{l}}$ not having any specific structure.

\subsection{Quantum circuit to generate the tensor network state \label{sec:QC}}

\begin{figure*}[tp]
\begin{center}
\includegraphics[scale=1.2]{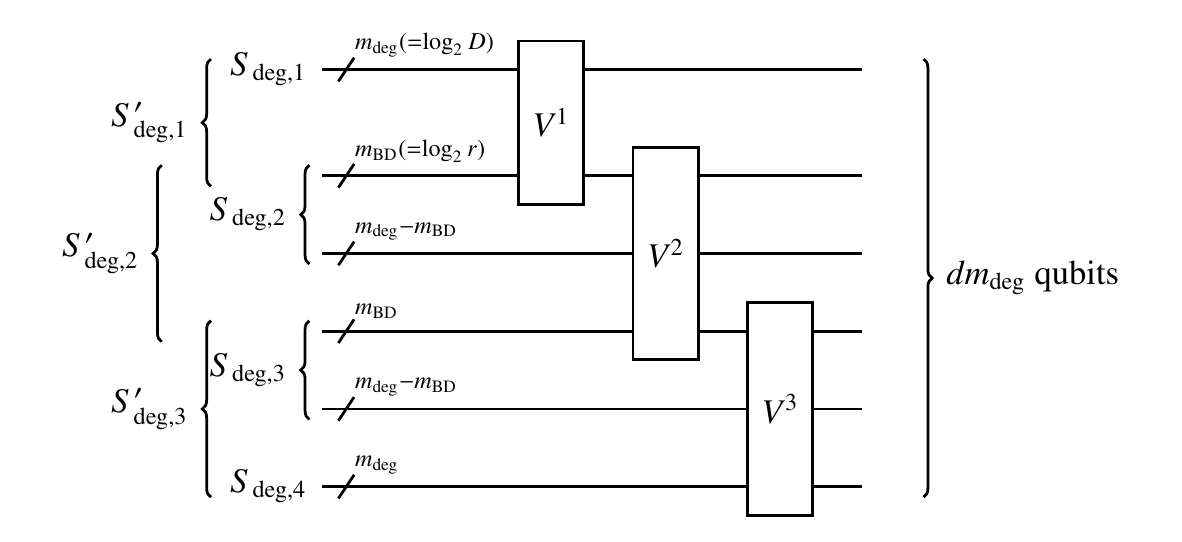}
\caption{Diagram of the quantum circuit $V_{\rm MPS}$ that generates the MPS-encoded state (\ref{eq:MPSState}) in the case of $d=4$. All the $dm_{\rm deg}$ qubits are initialized to $\ket{0}$. We describe the subsystems consisting of a portion of the qubits at the left end: $S_{{\rm deg},i}$ has $m_{\rm deg}$ qubits, the $((i-1)m_{\rm deg}+1)$-th to $im_{\rm deg}$-th ones, and $S_{{\rm deg},i}^\prime$ has $m_{\rm deg}+m_{\rm BD}$ qubits, the $((i-1)m_{\rm deg}+1)$-th to $(im_{\rm deg}+m_{\rm BD})$-th ones, except for $S_{{\rm deg},d-1}^\prime$ having the last $2m_{\rm deg}$ qubits.}
\label{fig:MPSCirc}
\end{center}
\end{figure*}

The quantum circuit to generate an MPS-encoded quantum state is shown in Fig. \ref{fig:MPSCirc}.
First, we prepare $dm_{\rm deg}$ qubits initialized to $\ket{0}$, where we assume that $D=2^{m_{\rm deg}}$ holds with some $m_{\rm deg}\in\mathbb{N}$.
Labeling them with the integers $1,...,dm_{\rm deg}$, for each $i\in[d]$, we denote the system of the $((i-1)m_{\rm deg}+1)$-th to $im_{\rm deg}$-th qubits by $S_{{\rm deg},i}$.
In addition, assuming that $r=2^{m_{\rm BD}}$ also holds with some $m_{\rm BD}\in\mathbb{N}$, for each $i\in[d-2]$, we denote the system of $S_{{\rm deg},i}$ and the first $m_{\rm BD}$ qubits in $S_{{\rm deg},i+1}$ by $S_{{\rm deg},i}^\prime$, and the system of the last $2m_{\rm deg}$ qubits by $S_{{\rm deg},d-1}^\prime$.
Then, we put the quantum gates $V_1,...,V_{d-1}$ on $S_{{\rm deg},1}^\prime,...,S_{{\rm deg},d-1}^\prime$, respectively, in this order.

Let us denote by $V_{\rm MPS}$ the unitary that corresponds to the whole of the above quantum circuit, which is depicted in Fig. \ref{fig:MPSCirc}. 
Then, $V_{\rm MPS}$ generates the MPS-encoded state
\begin{widetext}
\begin{equation}
    V_{\rm MPS}\ket{0}^{\otimes d}= \ket{\tilde{a}}:=\sum_{l_1=0}^{D-1}\cdots\sum_{l_{d}=0}^{D-1}\sum_{k_1=1}^{r}\cdots\sum_{k_{d-2}=1}^{r} U^{1}_{l_1,k_1} U^{2}_{k_1,l_2,k_2} \cdots U^{d-2}_{k_{d-3},l_{d-2},k_{d-2}} U^{d-1}_{k_{d-2},l_{d-1},l_d}\ket{l_1}\cdots\ket{l_d} \label{eq:MPSState},
\end{equation}
\end{widetext}
which is close to the state
\begin{equation}
    \ket{a}:=\sqrt{\frac{1}{\sum_{\vec{l}\in[D]_0^d} |a_{\vec{l}}|^2}}\sum_{\vec{l}\in[D]_0^d} a_{\vec{l}}\ket{l_1}\cdots\ket{l_d} \label{eq:coefstate}
\end{equation}
that encodes the true coefficients $a_{\vec{l}}$ if $\tilde{a}_{\vec{l}}$ in Eq. (\ref{eq:MPS}) approximates $a_{\vec{l}}$.
Here, $\ket{l}$ with $l\in[D]_0$ denotes the computational basis states on $S_{{\rm deg},1},...,S_{{\rm deg},d}$.
In addition, we associate the entries in $U^1,...,U^{d-1}$ with those in $V^1,...,V^{d-1}$ as unitaries, respectively, as follows:
\begin{equation}
    U^1_{l_1,k_1}=\bra{l_1+Dk_1}V^1\ket{0} \label{eq:UiQC1}
\end{equation}
for any $l_1\in[D]_0$ and $k_1\in[r]$,
\begin{equation}
    U^i_{k_{i-1},l_i,k_i}=\bra{l_i+Dk_i}V^i\ket{k_{i-1}} \label{eq:UiQC2}
\end{equation}
for any $i\in\{2,...,d-2\},l_i\in[D]_0$ and $k_{i-1},k_i\in[r]$, and
\begin{equation}
    U^{d-1}_{k_{d-2},l_{d-1},l_{d}}=\bra{l_{d-1}+Dl_{d}}V^{d-1}\ket{k_{d-2}} \label{eq:UiQC3}
\end{equation}
for any $l_{d-1},l_d\in[D]_0$ and $k_{d-2}\in[r]$, where $\ket{n}$ with $n\in\mathbb{N}_0$ denotes the computational basis state on either $S_{{\rm deg},1}^\prime,...,S_{{\rm deg},d-1}^\prime$.
Note that each $U^i$, which is not a unitary matrix, is realized as a block in $V^i$, a component unitary in the circuit $V_{\rm MPS}$ in Fig. \ref{fig:MPSCirc}.
Although $V^1,...,V^{d-1}$ have additional components that do not appear in Eqs. (\ref{eq:UiQC1}) to (\ref{eq:UiQC3}), those do not affect the state $\ket{\tilde{a}}$ because of the initialization of all qubits to $\ket{0}$.

Note that $U^2,...,U^{d-1}$ in Eqs. (\ref{eq:UiQC2}) and (\ref{eq:UiQC3}) automatically satisfy the conditions (\ref{eq:UiOrtho1}) and (\ref{eq:UiOrtho2}) because of the unitarity of $V^2,...,V^{d-1}$.
However, the unitarity of $V^1$ imposes the constraint
\begin{equation}
    \sum_{l_1=0}^{D-1}\sum_{k_1=1}^r |U^1_{l_1,k_1}|^2 = 1
\end{equation}
on $U^1$ in Eq. (\ref{eq:UiQC1}).
This implies that, with such $U^1$, the MPS-based approximation $\tilde{\tilde{f}}$ in Eq. (\ref{eq:MPSapp}) does not have the DOF of the overall factor, and therefore, although we can express the functional form of $f$ by $\tilde{\tilde{f}}$, we cannot adjust the magnitude of $\tilde{\tilde{f}}$ so that it fits $f$.
Conversely, if we have some estimate $C$ for the ratio of $f$ to $\tilde{\tilde{f}}$, we can approximate $f$ by $C\tilde{\tilde{f}}$.
This issue is addressed in Section \ref{sec:Opt}.


\subsection{Optimization of the quantum circuit \label{sec:Opt}}

We now consider how to optimize the quantum circuit $V_{\rm MPS}$ and obtain an MPS-based function approximation.

\begin{figure}[tp]
\begin{center}
\includegraphics[scale=1]{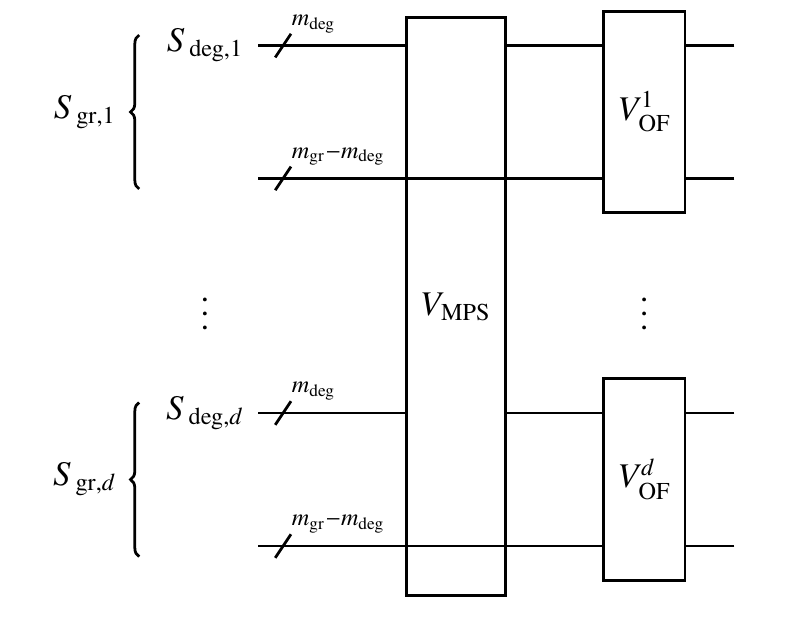}
\caption{Diagram of the quantum circuit $V_{\rm App}$ that generates the state (\ref{eq:MPSAppState}) in which the approximation $\tilde{\tilde{f}}$ in Eq. (\ref{eq:MPSapp}) of $f$ is amplitude-encoded.
$V_{\rm MPS}$ is the circuit depicted in Fig. \ref{fig:MPSCirc}; the lines that bypass it are not used in it.
All the $dm_{\rm gr}$ qubits are initialized to $\ket{0}$.
The subsystems consisting of a portion of the qubits, which are indicated at the left end, are described in the body text.}
\label{fig:MPSAppCirc}
\end{center}
\end{figure}

First, we extend the circuit $V_{\rm MPS}$ in Section \ref{sec:QC} using $\{V^i_{\rm OF}\}_i$ as follows.
For each $i\in[d]$, we add $m_{\rm gr}-m_{\rm deg}$ qubits after the $im_{\rm deg}$-th qubits in the original circuit and denote the system consisting of $S_{{\rm deg},i}$ and the added qubits by $S_{{\rm gr},i}$.
Note that the resultant system is the same as the system $S$ for $O_f$, which consists of the $d$ $m_{\rm gr}$-qubit registers.
We then perform $V^i_{\rm OF}$ on $S_{{\rm deg},i}$.
The resultant circuit $V_{\rm App}$ is shown in Fig. \ref{fig:MPSAppCirc}.
Note that this circuit generates
\begin{widetext}
\begin{equation}
    V_{\rm App}\ket{0}^{\otimes d}=\Ket{\tilde{\tilde{f}}}:= \sum_{l_1=0}^{D-1}\cdots\sum_{l_{d}=0}^{D-1}\sum_{k_1=1}^{r}\cdots\sum_{k_{d-2}=1}^{r} \sum_{j_1=0}^{n_{\rm gr}-1}\cdots\sum_{j_d=0}^{n_{\rm gr}-1}U^{1}_{l_1,k_1} U^{2}_{k_1,l_2,k_2} \cdots U^{d-2}_{k_{d-3},l_{d-2},k_{d-2}} U^{d-1}_{k_{d-2},l_{d-1},l_d}P^1_{l_1}(x_{1,j_1})\cdots P^d_{l_d}(x_{d,j_d})\ket{j_1}\cdots\ket{j_d}, \label{eq:MPSAppState}
\end{equation}
\end{widetext}
where $\ket{n}$ with $n\in\mathbb{N}_0$ now denotes the computational basis state on $S_{{\rm gr},1},...,S_{{\rm gr},d}$.
That is, this is the quantum state that amplitude-encodes $\tilde{\tilde{f}}$ in Eq. (\ref{eq:MPSapp}), the approximation of $f$ by the orthogonal function expansion and the MPS approximation of the coefficients.
Therefore, if we obtain $V_{\rm App}$ that generates $\Ket{\tilde{\tilde{f}}}$ close to $\Ket{f}$, we also obtain $\{U^i\}_i$ for which $\tilde{\tilde{f}}$ in Eq. (\ref{eq:MPSapp}) well approximate $f$ at least on the grid points, by reading out their entries from the quantum gates $\{V^i\}_i$ in $V_{\rm App}$, except for the overall factor $C$.

Next, we consider how to obtain such $V_{\rm App}$, especially $\{V^i\}_i$ in it.
We aim to maximize the fidelity
\begin{equation}
    F=\Braket{f | \tilde{\tilde{f}}}. \label{eq:fid}
\end{equation}
Note that 
maximizing $F$ is equivalent to minimizing the sum of the squared differences between the two normalized functions 
\begin{equation}
    \sum_{\vec{j}\in[n_{\rm gr}]_0^d} \left(\frac{f(\vec{x}_{\vec{j}})}{C}-\tilde{\tilde{f}}(\vec{x}_{\vec{j}})\right)^2,
\end{equation}
which, in the large $n_{\rm gr}$ limit, is equivalent to
\begin{equation}
    \int_{\Omega}  \left(\frac{f(\vec{x}_{\vec{j}})}{C}-\tilde{\tilde{f}}(\vec{x}_{\vec{j}})\right)^2 w^1(x_1)\cdots w^d(x_d) d\vec{x},
\end{equation}
that is, the squared $L^2$ norm of $\frac{f}{C}-\tilde{\tilde{f}}$, the common metric in function approximation.

The procedure for maximizing $F$ is similar to that presented in \cite{shirakawa2021}.
We attempt to optimize each $V^i$ alternatingly.
In other words, we optimize each $V^i$ with the others fixed, setting $i$ to $1,2,...,{d-1}$ in turn.
This loop can be repeated an arbitrary number of times.
The optimization step for $V^i$ proceeds as follows.
We define
\begin{equation}
    F_i = {\rm Tr}_{\bar{S}^\prime_{{\rm deg},i}}  \left[ \ket{\Psi_{i+1}} \bra{\Phi_{i-1}} \right].
\end{equation}
Here, $\bar{S}^\prime_{{\rm deg},i}$ denotes the system consisting of the qubits except those in $S^\prime_{{\rm deg},i}$, and, for any subsystem $s$ in $S$, ${\rm Tr}_{s}$ denotes the partial trace over the Hilbert space corresponding to $s$.
$\ket{\Psi_{i+1}}$ is defined as
\begin{eqnarray}
    &&\ket{\Psi_{i+1}} = \nonumber \\
    && \begin{cases}
    \left(V^{i+1} \otimes I_{\bar{S}^\prime_{{\rm deg},i+1}}\right)^\dagger\cdots\left(V^{d-1} \otimes I_{\bar{S}^\prime_{{\rm deg},d-1}}\right)^\dagger (V^1_{\rm OF}\otimes\cdots\otimes V^{d-1}_{\rm OF})^\dagger \ket{f} \\
    \qquad\qquad\qquad\qquad\qquad\qquad\qquad\qquad ; \ {\rm for} \ i=1,...,d-2 \\
    (V^1_{\rm OF}\otimes\cdots\otimes V^{d-1}_{\rm OF})^\dagger\ket{f} \ \ \ \qquad\qquad\qquad ; \ {\rm for} \ i=d-1
    \end{cases} \nonumber \\
    && 
\end{eqnarray}
where $I_{\bar{S}^\prime_{{\rm deg},i}}$ denotes the identity operator on $\bar{S}^\prime_{{\rm deg},i}$, and thus $V^{i} \otimes I_{\bar{S}^\prime_{{\rm deg},i}}$ denotes the $i$-th block in $V_{\rm MPS}$ in Fig. \ref{fig:MPSCirc}.
$\ket{\Phi_{i-1}}$ is defined as
\begin{equation}
    \ket{\Phi_{i-1}} =
    \begin{cases}
    \left(V^{i-1} \otimes I_{\bar{S}^\prime_{{\rm deg},i-1}} \right)\cdots\left(V^{1} \otimes I_{\bar{S}^\prime_{{\rm deg},1}} \right)\ket{0}^{\otimes d}  \\
    \qquad\qquad\qquad\qquad\qquad\qquad ; \ {\rm for} \ i=2,...,d-1 \\
    \ket{0}^{\otimes d} \qquad\qquad\qquad\qquad \ \ \ \ ; \ {\rm for} \ i=1
    \end{cases}
\end{equation}
We can regard this $F_i$ as an $M\times M$ matrix, where $M=rD$ for $i\in[d-2]$ and $M=D^2$ for $i=d-1$.
Its entries are given by $\bra{l}F_i\ket{l^\prime}$, where $\ket{l},\ket{l^\prime}\in\{\ket{0},\ket{1},...,\ket{M-1}\}$ are now the computational basis states on $S^\prime_{{\rm deg},i}$.
Supposing that we know $F_i$, we perform its singular value decomposition (SVD)
\begin{equation}
F_i=XDY, \label{eq:FiSVD}
\end{equation}
where $X$ and $Y$ are the $M\times M$ unitaries and $D$ is the diagonal matrix having the singular values of $F_i$ as its diagonal entries.
Finally, we update $V^i$ by
\begin{equation}
    V^i = XY. \label{eq:ViUpdate}
\end{equation}

We obtain $F_i$ through Pauli decomposition and the Hadamard test \cite{shirakawa2021}.
We set $m=\log_2 M$, and, for any $k\in[m]$, we denote the identity operator, the Pauli-X gate, the Pauli-Y gate, and the Pauli-Z gate on the $k$-th qubit in $S^\prime_{{\rm deg},i}$ by $\hat{\sigma}^k_0,\hat{\sigma}^k_1,\hat{\sigma}^k_2$ and $\hat{\sigma}^k_3$, respectively.
We also define
\begin{equation}
    \hat{\sigma}_{\vec{\alpha}}:= \hat{\sigma}^1_{\alpha_1}\otimes\cdots\otimes\hat{\sigma}^{m}_{\alpha_{m}}
\end{equation}
for any $\vec{\alpha}=(\alpha_1,...,\alpha_{m})\in\{0,1,2,3\}^{m}$.
Then, we can always decompose $F_i$ as
\begin{equation}
    F_i = \sum_{\vec{\alpha}\in\mathcal{A}}\tilde{F}_{i,\vec{\alpha}} \hat{\sigma}_{\vec{\alpha}}, \label{eq:FiPauli}
\end{equation}
with $\tilde{F}_{i,\vec{\alpha}}\in\mathbb{R}$, where, rather than $\{0,1,2,3\}^{m}$, the index tuple $\vec{\alpha}$ runs over the subset
\begin{equation}
    \mathcal{A}:=\left\{\vec{\alpha}\in\{0,1,2,3\}^{m} \ \middle| \ {\rm the \ number \ of \ 2's \ in} \ \alpha_1,...,\alpha_m \ {\rm is \ even.} \right\}
\end{equation}
because $F_i$ is real.
Because
\begin{equation}
    {\rm Tr}_{S^\prime_{{\rm deg},i}}  \left[ F_i \hat{\sigma}_{\vec{\alpha}}\right]=M\tilde{F}_{i,\vec{\alpha}}
\end{equation}
holds, we obtain $\tilde{F}_{i,\vec{\alpha}}$ by estimating ${\rm Tr}_{S^\prime_{{\rm deg},i}}  \left[ F_i \hat{\sigma}_{\vec{\alpha}}\right]$.
Noting that
\begin{equation}
    {\rm Tr}_{S^\prime_{{\rm deg},i}}  \left[ F_i \hat{\sigma}_{\vec{\alpha}}\right] = \bra{\Phi_{i-1}} \hat{\sigma}_{\vec{\alpha}} \ket{\Psi_{i+1}}, \label{eq:TrFsigma}
\end{equation}
we can estimate this using the Hadamard test, in which the circuit $V_{\rm MPS}$ with replacement of $V^i$ by $\hat{\sigma}_{\vec{\alpha}}$ is used.
In one update of each $V^i$, the total number of estimations of the quantities (\ref{eq:TrFsigma}) is $|\mathcal{A}|=M(M+1)/2$.
We refer to \cite{shirakawa2021} for the details of the estimations.

After we optimize the $\{V^i\}_i$ and read out $\{U^i\}_i$ from them, the remaining task is just multiplying the factor $C$ to $\tilde{\tilde{f}}$ constructed from $\{U^i\}_i$.
In this paper, we do not go into the details of estimating this factor but simply assume that we have some estimate for it.
In some quantum algorithms for solving PDEs, the method of estimating this factor, the root of the squared sum of the function values on the grid points, is presented \cite{linden2020quantum}.

The entire procedure to obtain an approximation of $f$ is summarized as Algorithm \ref{alg:proposed}.

\begin{figure}[htp]
\begin{algorithm}[H]
	\caption{Proposed algorithm to obtain an approximation of the function $f:\Omega\rightarrow\mathbb{R}$.} 
	\label{alg:proposed}
	\begin{algorithmic}[1]
		\REQUIRE{\ \\
        \begin{itemize}
        \item $D=2^{m_{\rm deg}}$ with $m_{\rm deg}\in\mathbb{N}$: the degree of the orthogonal functions
        \item $r=2^{m_{\rm BD}}$ with $m_{\rm BD}\in\mathbb{N}$: the bond dimension
        \item The oracle $O_f$ in Eq. (\ref{eq:Of}). 
        \item The orthogonal functions $\{P^i_l\}_{i\in[d],l\in[D]_0}$ satisfying Eq. (\ref{eq:orthoDisc}).
        \item The iteration $n_{\rm iter}\in\mathbb{N}$ of the optimization loop.
        \item The initial values of the $rD\times rD$ unitaries $V^1,...,V^{d-2}$ and the $D^2\times D^2$ unitary $V^{d-1}$, which may be chosen randomly.
        \item $C$ in Eq. (\ref{eq:C}). 
        \end{itemize}
		} 		

        \FOR{$i_{\rm iter}=1$ to $n_{\rm iter}$}
            \FOR{$i=1$ to $d-1$}
                \FOR{$\vec{\alpha}\in\{0,1,2,3\}^{m}$}
                    \STATE Estimate $\bra{\Phi_{i-1}} \hat{\sigma}_{\vec{\alpha}} \ket{\Psi_{i+1}}$ using the Hadamard test, and divide it by $M$ to obtain $\tilde{F}_{\vec{\alpha}}$.
                \ENDFOR
                \STATE Set $F_i$ as per Eq. (\ref{eq:FiPauli}).
                \STATE Perform the SVD of $F_i$ as per Eq. (\ref{eq:FiSVD}).
                \STATE Update $V^i$ as per Eq. (\ref{eq:ViUpdate}).
            \ENDFOR
        \ENDFOR

        \STATE Set $U^1\in\mathbb{R}^{D \times r}$, $U^i\in\mathbb{R}^{r\times D \times r}$ for $i\in\{2,...,d-2\}$, and $U^{d-1}\in\mathbb{R}^{r \times D\times D}$ as per Eqs. (\ref{eq:UiQC1}), (\ref{eq:UiQC2}) and (\ref{eq:UiQC3}), respectively.

        \STATE Define $\tilde{\tilde{f}}$ as per Eq. (\ref{eq:MPSapp}).

        \STATE Output $C\tilde{\tilde{f}}$ as an approximation of $f$.
	\end{algorithmic}
\end{algorithm}
\end{figure}

\subsection{Situation in which our method is useful}

In the introduction, we mentioned solving PDEs using quantum algorithms as a main use case of our method.
One might think that if the solution of the PDE can be approximated by MPS as Eq. (\ref{eq:tiltilf}), we can directly plug the ansatz (\ref{eq:tiltilf}) into the PDE and optimize the parameters $\{U^i_{k_{i-1},l_i,k_i}\}$ using a classical computer.
In fact, such a method has been studied in \cite{BOELENS2018519,DEKTOR2020109125,DEKTOR2021110295,dektor2021rank}.
However, the scheme we are now considering, that is, generating the quantum state that encodes the solution in the amplitude by a quantum algorithm and reading out the solution in MPS-based approximation, is applicable to the broader range of problems.
For example, in the initial value problem, it is possible that MPS-based approximation is valid at the terminal time but not at some intermediate time points.
In the context of quantum many-body physics, this is applicable to the wave function of a system that changes to a low-entangled state via a highly entangled state, such as quantum annealing~\cite{QA,QA_review_1,QA_review_2}.
In such a case, finding the function at the terminal time using a quantum PDE solver and then reading it out by MPS-based approximation might work, although using MPS throughout does not seem to work.

\section{Numerical experiment \label{sec:NumExp}}

We now confirm the feasibility of our method through a numerical experiment on approximating a finance-motivated function.

\subsection{Problem setting \label{sec:AppTgt}}

As a reasonable instance of the target function for the approximation, we take the price of a {\it financial derivative} (simply, {\it derivative}).

Specifically, we consider the worst-of put option written on the $d$ underlying assets that obey the Black-Scholes (BS) model and take its present price as a function $f(\vec{s})$ of the asset prices $\vec{s}=(s_1,...,s_d)$ as the approximation target.
This is the solution of the so-called Black-Scholes PDE and thus fits the current situation in which a quantum PDE solver outputs the state $\ket{f(\vec{s})}$. 
Details are provided in Appendix \ref{sec:deriv}.

We approximated $f(\vec{s})$ on the hyperrectangle $[L_1,U_1]\times\cdots\times[L_d,U_d]$.
We took cosine functions as orthogonal functions.
Specifically, for any $i\in[d]$ and $l\in[D]_0$, we set
\begin{equation}
    P^i_{l}(s_i) =
    \cos\left(l\frac{s_i-L_i}{U_i-L_i}\pi\right). \label{eq:cos}
\end{equation}
The settings for $U_i$ and $L_i$ are presented in Appendix \ref{sec:deriv}.
The orthogonal relation (\ref{eq:orthoDisc}) is satisfied with the grid points set as
\begin{equation}
    s_{i,j}=\frac{j+\frac{1}{2}}{n_{\rm gr}}(U_i-L_i)+L_i
\end{equation}
for $j\in[n_{\rm gr}]_0$ and $c^i_l$ being
\begin{equation}
    c^i_l=
    \begin{cases}
        n_{\rm gr} & ; \ {\rm for} \ l=0 \\
        \frac{n_{\rm gr}}{2} & ; \ {\rm for} \ l\in[D-1] 
    \end{cases}.
\end{equation}

Let us comment on the choice of cosines as orthogonal functions.
We consider the case where the grid points are equally spaced points in the hyperrectangle, which we expect to be the simplest and most common.
In this setting, the cosine functions (\ref{eq:cos}) satisfy the orthogonal relation (\ref{eq:ortho}) exactly.
Therefore, the choice was natural in this case.
Generally, it is plausible to take orthogonal functions such that the orthogonal relation holds for the given grid points.

\subsection{Algorithm modifications to run the numerical experiment}

We used Algorithm \ref{alg:proposed} but made the following modification, since we performed all calculations on a classical computer, and thus there was a memory space limitation.
Rather than $F$ in Eq. (\ref{eq:fid}), we attempt to maximize
\begin{equation}
    F^\prime = \Braket{a | \tilde{a}}
\end{equation}
Here, $\ket{\tilde{a}}$ is given in Eq. (\ref{eq:MPSState}) and
\begin{equation}
    \ket{a}:=\sqrt{\frac{1}{\sum_{\vec{l}\in[D]_0^d} |a_{\vec{l}}|^2}}\sum_{\vec{l}\in[D]_0^d} a_{\vec{l}}\ket{l_1}\cdots\ket{l_d},
\end{equation}
where we calculated the coefficients $a_{\vec{l}}$ for each $\vec{l}\in[D]_0^d$ using Eq. (\ref{eq:coef}) and the values of $f(\vec{s})$ on the grid points computed by Monte Carlo integration (see Appendix \ref{sec:deriv}). 
We take the minimal grid points to calculate the coefficients for a given $D$, which means $n_{\rm gr}=D$, although it is assumed that $n_{\rm gr}>D$ when our algorithm is used on a future quantum computer with the oracle $O_f$.

With this modification, the quantum circuit under consideration becomes small, and the calculation becomes feasible on a classical computer.
The gates $\{V^i_{\rm OF}\}$ do not appear in our experiment, and their roles are absorbed into the aforementioned preprocessing to calculate $\{a_{\vec{l}}\}$.
Most importantly, note that $\{U^i\}$ calculated under the above modification are the same as those output by our algorithm without modification.

\subsection{Result of the approximation}

Next, we attempted to obtain an approximation of $f(\vec{s})$.
We set the parameters as follows: $d=5,D=r=16,n_{\rm iter}=5$.
For $C$, we used the root squared sum of the values of $f(\vec{s})$ on the grid points, which are calculated using Monte Carlo integration.
We used the ITenosr library \cite{itensor} for the tensor calculations.

\begin{figure}[tp]
		\centering
		\includegraphics[scale=1.0]{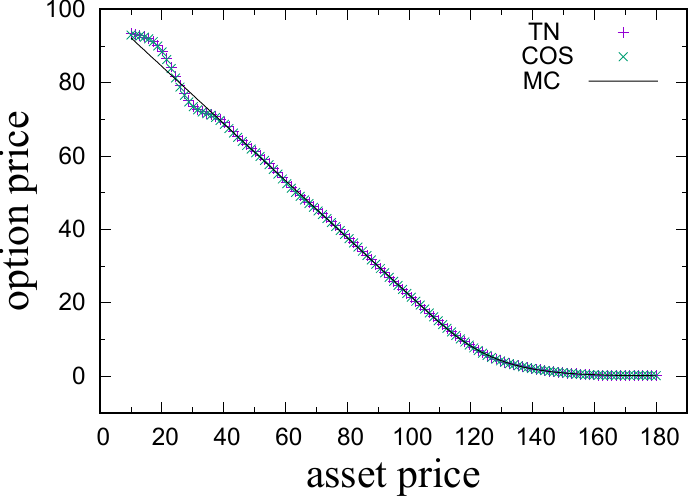}
		\caption{Five-asset worst-of put option prices $f(\vec{s})$ for $s_1=...=s_5=s$ calculated in three different ways. The horizontal and vertical axes, respectively, correspond to $s$ and $f(\vec{s})$. The $+$, $\times$, and black line, 
		respectively, indicate results obtained from MPS-based approximation, cosine expansion approximation, and Monte Carlo integration.}
	\label{fig:res}
\end{figure}

We now demonstrate the accuracy of the obtained MPS-based approximation of $f(\vec{s})$.
However, because we are considering a high-dimensional space, it is difficult to display the accuracy over the entire space.
Therefore, we show the accuracy on the following two sets of selected points: (i) the ``diagonal" line $s_1=...=s_d$ and (ii) $10^4$ random points sampled according to the BS model (see Appendix \ref{sec:deriv} for details).

Fig. \ref{fig:res} displays results for the diagonal line, depicting the MPS-based approximation $\tilde{\tilde{f}}(\vec{s})$ of $f(\vec{s})$ along with Monte Carlo integration values and the cosine expansion approximation as per Eq. (\ref{eq:orthoExp}).
The MPS-based approximation fits the Monte Carlo integration values well within an error of about 0.5 or less, except for the region near the lower end of the domain of the approximations.
Note that the cosine expansion approximation already has an error from the Monte Carlo values, and this acts as a lower bound on the error of the MPS-based approximation.
The MPS-based approximation almost overlaps the cosine expansion approximation, which means that the MPS-based approximation worked well.

The results for the random sample points are summarized in Table \ref{tab:resRand}.
On these points, the maximum differences among the values of $f(\vec{s})$ calculated by Monte Carlo integration, MPS-based approximation, and cosine expansion using the full coefficients without MPS approximation are about 0.5 or less, similar to most regions in the diagonal line case.
This result supports the proposed method.

\begin{table}[tp]
\centering
 \caption{Maximum differences among the values of the five-asset worst-of put option prices $f(\vec{s})$ calculated by Monte Carlo integration (MC), MPS-based approximation (TN), and cosine expansion using the full coefficients (COS), on the random sample points.}
 \label{tab:resRand}
  \begin{tabular}{cc}
   \hline
   $\max\left|{\rm TN} \ - {\rm MC}\right|$ & 0.5086 \\
   $\max\left|{\rm COS} \ - {\rm MC}\right|$ & 0.5102 \\
   $\max\left|{\rm TN} \ - {\rm COS}\right|$ & 0.3164 \\
   \hline
  \end{tabular}
\end{table}

In the current MPS-based approximation, the number of DOF is 12544, which is smaller than the number of cosine expansion coefficients (1048576) by two orders of magnitude.
Therefore, we achieved a large parameter reduction while maintaining the approximation accuracy.

\subsection{Relationship between approximation accuracy and degrees of freedom}

\begin{figure*}[tp]
\begin{center}
\includegraphics[scale=1]{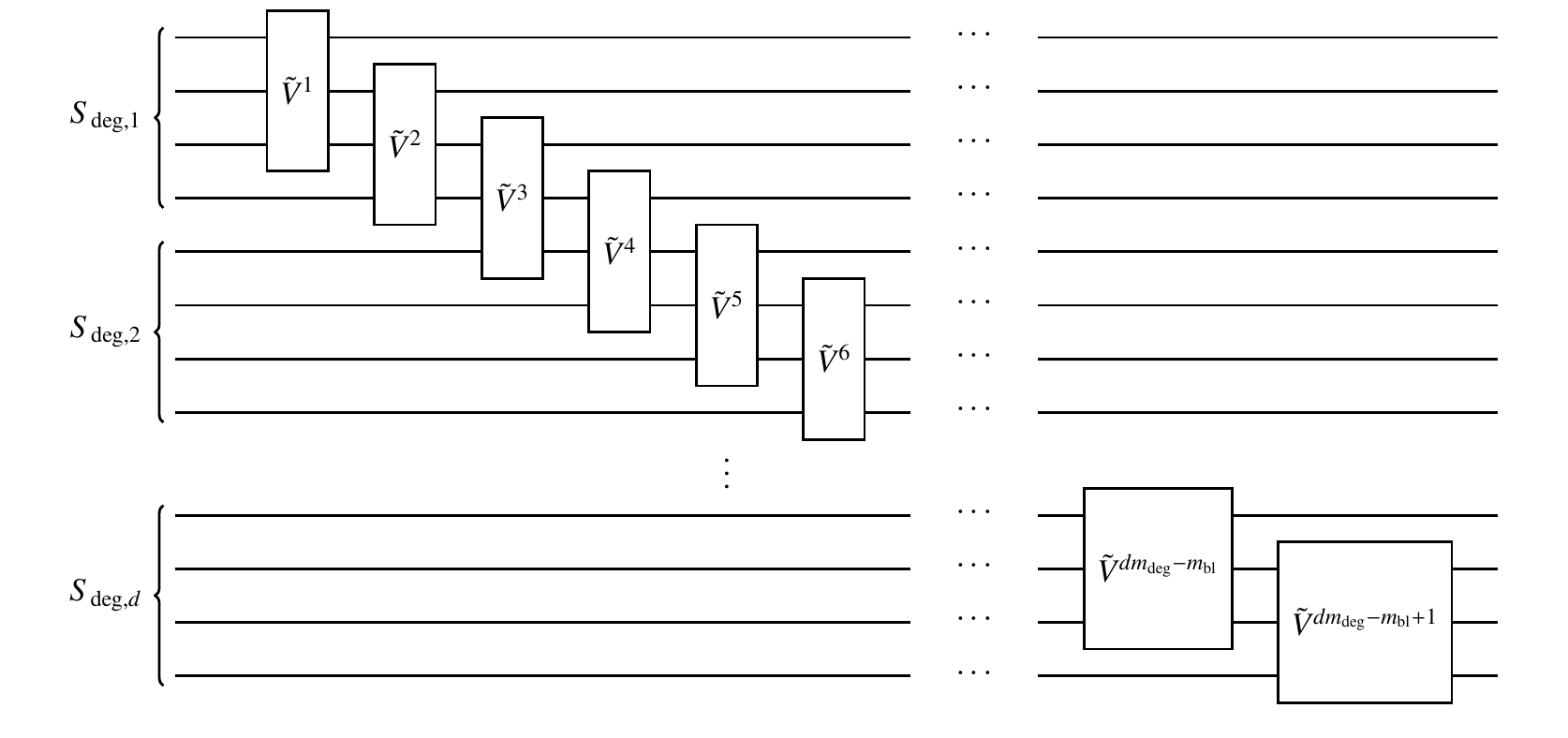}
\caption{Diagram of the quantum circuit $V^\prime_{\rm MPS}$ used in the additional numerical experiment instead of $V_{\rm MPS}$ for $m_{\rm deg}=4$ and $m_{\rm bl}=3$.}
\label{fig:TestCirc}
\end{center}
\end{figure*}

To investigate the relationship between the DOF and the approximation accuracy, we performed the following additional experiment.
We replaced the circuit $V_{\rm MPS}$ in Fig. \ref{fig:MPSCirc} by a different $V_{\rm MPS}^\prime$ having the different DOF.
Here, with $m_{\rm bl}\in\{2,...,dm_{\rm deg}-1\}$, $V_{\rm MPS}^\prime$ consists of the gates $\tilde{V}^1,...,\tilde{V}^{dm_{\rm deg}-m_{\rm bl}+1}$ that act on the systems of $m_{\rm bl}$ qubits displaced one by one, as shown in Fig. \ref{fig:TestCirc}.
The DOF of this circuit is
\begin{equation}
    2^{m_{\rm bl}}+2^{2m_{\rm bl}-1}(dm_{\rm deg}-m_{\rm bl}).
\end{equation}
We alternatingly optimized the blocks $\{\tilde{V}^i\}$ in a manner similar to Algorithm \ref{alg:proposed} so that $\braket{a | \tilde{a}^\prime}$ is maximized, where 
\begin{equation}
\ket{\tilde{a}^\prime}:=\sum_{\vec{l}\in[D]_0^d}\tilde{a}^\prime_{\vec{l}}\ket{l_1}\cdots\ket{l_d}
\end{equation}
is the state generated by $V_{\rm MPS}^\prime$.
Then, we obtained the approximation of $f(\vec{s})$ as
\begin{equation}
 \tilde{\tilde{f}}^\prime(\vec{s}):=C\sum_{\vec{l}\in[D]_0^d}\tilde{a}^\prime_{\vec{l}}P_{\vec{l}}(\vec{s})~.    
\end{equation}

\begin{figure}[tp]
    \centering
    \subfigure[On the diagonal line. $a=e^{7.27}$ and $b=-0.87$.]{
    \includegraphics[scale=1.0]{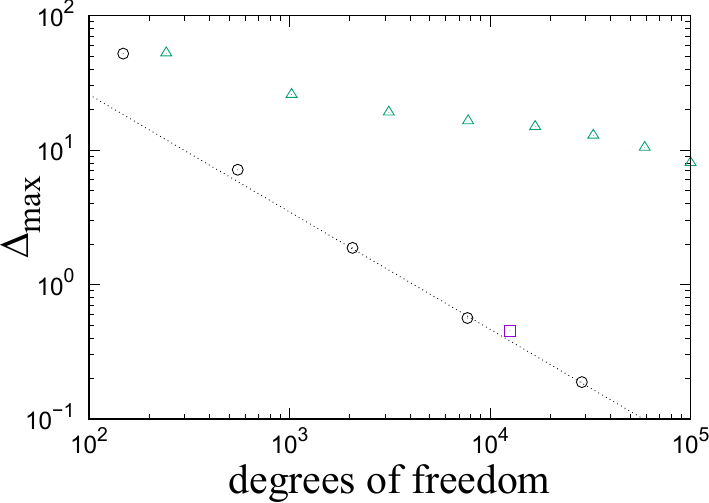} 	\label{fig:dof}
    }
    \subfigure[On the random sample points. $a=e^{8.75}$ and $b=-1.07$.]{
    \includegraphics[scale=1.0]{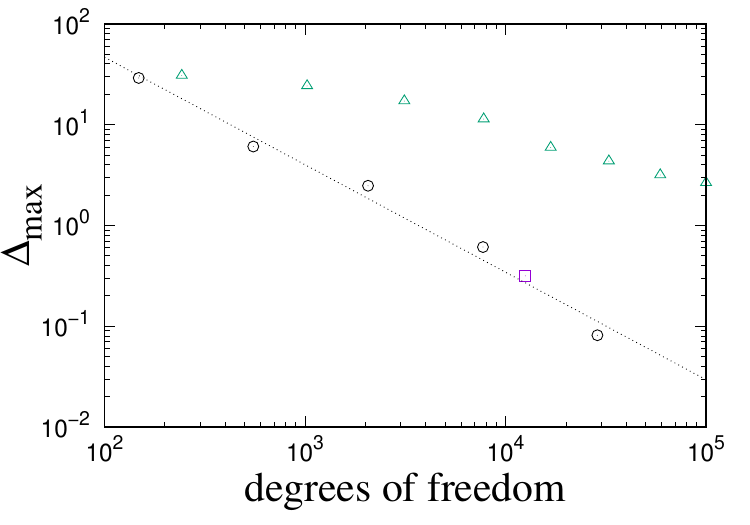} \label{fig:dof_rand}
    }
    \caption{Maximum difference $\Delta_{\rm max}$ of the various approximations for the five-asset worst-of put option price from the cosine expansion approximation of degree $D=16$. 
    The circles indicate the approximations based on the circuit $V^\prime_{\rm MPS}$ in Fig. \ref{fig:TestCirc}.
    From left to right, the points correspond to $m_{\rm bl}=2,...,6$, respectively.
    The dotted lines represent the function $\Delta_{\rm max}(x)=ax^b$ of the degrees of freedom $x$ with $a$ and $b$ fitted with respect to the data for $m_{\rm bl}=[4,6]$ for (a) and $m_{\rm bl}=[2,6]$ for (b).
    The square indicates the maximum difference of the MPS-based approximation, and the triangles indicate those of the cosine expansion approximations of degree $D=3,...,10$, which correspond to the points from left to right, respectively.
    The vertical and horizontal lines correspond to the maximum difference and DOF of the approximations, respectively.}
\end{figure}

In Fig. \ref{fig:dof}, we display the maximum difference of $\tilde{\tilde{f}}^\prime$ from the cosine expansion approximation on the diagonal line $s_1=...=s_d$, taking $m_{\rm bl}=2,...,6$, along with that of the MPS-based approximation.
This figure shows that there is a power law relationship between the approximation accuracy and DOF in the region with large DOF. 
This behavior is similar to that often observed in critical MPS systems applied to one-dimensional quantum systems or two-dimensional classical systems\cite{NISHINO199669,PhysRevB.78.024410,PhysRevLett.102.255701,PhysRevB.86.075117,PhysRevB.87.235106,PhysRevB.91.035120,PhysRevE.96.062112}.
If this behavior also appears in the case of $d \gg 1$, one might make a similar argument to the finite bond-dimension (entanglement) scaling refined in the study of MPS and evaluate $\tilde{f}$ with appropriate extrapolations with respect to DOF.

We then show the relationship between the accuracy of the cosine expansion approximation and its DOF in Fig. \ref{fig:dof}.
We plot the maximum error of the approximated worst-of put option price by the cosine expansions of low degree $D=3,...,10$ on the line $s_1=...=s_d$, taking that of degree $D=16$ as the reference value.
The error of the low-degree cosine expansion is much larger than that of $\tilde{\tilde{f}}^\prime$ with comparable DOF.
Fig. \ref{fig:dof_rand} is similar to Fig. \ref{fig:dof} but for the random sample points.
Again, the maximum difference of $\tilde{\tilde{f}}^\prime$ from the cosine expansion with $D=16$ displays a power law with respect to DOF and is much smaller than the error of the low-degree cosine expansion.

These results provide additional evidence of the advantages of our MPS-based approximation and the approximation by the circuit $V^\prime_{\rm MPS}$ over the simple cosine expansion.

\section{Summary \label{sec:sum}}

In this study, we have considered how to extract the function encoded in the amplitudes of the quantum state as classical data.
Such a task necessarily accompanies quantum algorithms such as PDE solvers, but to date there has not been a proposal for a general method to accomplish this task, even though it risks ruining quantum speedup.
We proposed a method based on orthogonal function expansion and tensor network.
Orthogonal function expansion is widely used for function approximation but suffers from an exponential increase in the number of the parameters, the expansion coefficients, with respect to the dimension, that is, the number of variables of the function.
We then use an MPS, a type of tensor network, to approximate the coefficients as a high-order tensor and reduce the DOF.
We presented a quantum circuit that produces the state corresponding to such a function approximation.
Such a circuit is, in fact, constructible because an MPS is encoded in a quantum state by a simple circuit, and so are orthogonal functions because of their orthogonal relation.
We also presented the procedure to optimize the quantum circuit and MPS-based approximation, based on the alternating method proposed in \cite{shirakawa2021}.
Finally, we conducted a numerical experiment to approximate a finance-motivated multivariate function and found that our method works in this case.

This study has scope for further investigation.
For example, we did not explicitly present the bound on the total query complexity for the oracles $O_f$ and $V^i_{\rm OF}$ to obtain an approximating function with a given accuracy $\epsilon$.
Presenting such a complexity estimation is desired but difficult at present.
In general, function approximation in a high-dimensional space is a long-standing problem in computational mathematics.
Although a recent study showed that there is an MPS-based approximation that achieves a given accuracy with DOF subexponential with respect to $d$ for any function with some property \cite{griebel2022low}, to the best of the authors' knowledge, there is no known algorithm that certainly outputs such an approximation.
The convergence
property of the alternating optimization method explained above is not well known theoretically, even though its good performance is empirically known.
In these situations, we have considered a quantitative discussion on the accuracy and complexity of our method beyond the scope of this study and left it for future work, presenting some numerical evidence instead.

We expect that the proposed method can be used widely in combination with various quantum algorithms that output function-encoding states, whether for FTQC or NISQ.
In future work, we will attempt to combine this method with concrete function-finding quantum algorithms on concrete problems and present a complete set of the algorithms, along with a more quantitative analysis of complexity. 

\section*{acknowledgement}

This work was partially supported by
KAKENHI Grant Numbers JP22K11924, JP21H04446, JP21H05182, and JP21H05191 from JSPS of Japan and also by JST PRESTO No. JPMJPR1911, MEXT Q-LEAP Grant No. 
JPMXS0120319794, and JST COI-NEXT No. JPMJPF2014.
H.U was supported by a COE research grant in computational science from Hyogo Prefecture and Kobe City through the Foundation for Computational Science.

\section*{Data availability}

The datasets generated during and/or analyzed during the current study are available from the corresponding author on reasonable request.

\section*{Conflict of interest}

The authors declare no conflict of interest.

\appendix

\section{How to construct $V^i_{\rm OF}$ \label{sec:ViOF}}

\begin{figure}[t]
\begin{minipage}[b]{\linewidth}
\begin{center}
\includegraphics[scale=1.2]{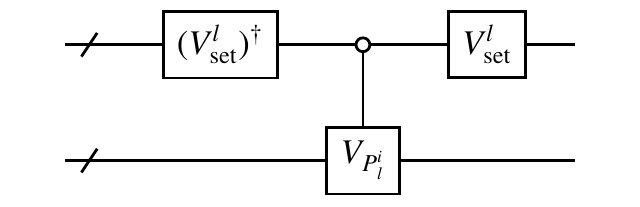}
    \caption{Implementation of $V_{P^i_l}^\prime$ in Eq. (\ref{eq:VPilPr}). The open circle represents the control on the gate $V_{P^i_l}$ such that it is activated if and only if all the qubits in the first register take $\ket{0}$.}
    \label{fig:VPilPr}
    \end{center}
\end{minipage}

\begin{minipage}[b]{\linewidth}
\begin{center}
\includegraphics[scale=1.2]{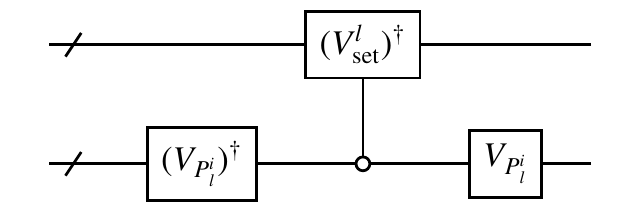}
    \caption{Implementation of $V^{i,l}_{\rm reset}$ in Eq. (\ref{eq:reset}).}
    \label{fig:reset}
    \end{center}
\end{minipage}
\end{figure}

To construct $V^i_{\rm OF}$ as per Eq. (\ref{eq:ViOF}), we require some building block quantum gates.
First, for any $i\in[d]$ and $l\in[D]_0$, we assume the availability of the gate $V_{P^i_l}$ that acts on an $m_{\rm gr}$-qubit register as
\begin{equation}
    V_{P^i_l}\ket{0} = \ket{P^i_l}, \label{eq:VPil}
\end{equation}
where $\ket{P^i_l}$ is given in Eq. (\ref{eq:ViOF}).
In fact, commonly used orthogonal functions such as trigonometric functions and orthogonal polynomials are explicitly given as elementary functions, and thus the state generation oracle as in Eq. (\ref{eq:VPil}) is constructed using the Grover-Rudolph method \cite{grover2002creating,kaneko2022quantum}.
Second, for any  $n\in\mathbb{N}_0$, we denote by $V^n_{\rm set}$ the gate that acts on a quantum register with a sufficient number of qubits as
\begin{equation}
    V^n_{\rm set}\ket{0}=\ket{n}.
\end{equation}
This can be constructed by putting a Pauli-X gate (resp. nothing) on the $a$-th qubit in the register if the binary representation of $n$ contains a 1 (resp. 0) in the $a$-th digit.

Then, for any $i\in[d]$ and $l\in[D]_0$, using the controlled $V_{P^i_l}$ and $V^l_{\rm set}$, we can implement the following gate on the system of two $m_{\rm gr}$-qubit registers
\begin{equation}
    V_{P^i_l}^\prime := \ket{l}\bra{l} \otimes V_{P^i_l} + \sum_{l^\prime\in[n_{\rm gr}]_0\setminus\{l\}} \ket{l^\prime}\bra{l^\prime} \otimes I, \label{eq:VPilPr}
\end{equation}
where $I$ is the identity operator, as shown in Fig. \ref{fig:VPilPr}.
Combining gates of this type, we obtain
\begin{equation}
    V^{i}_{\rm OF 1} := \prod_{l=0}^{D-1}V_{P^i_l}^\prime=\sum_{l=0}^{D-1} \ket{l}\bra{l} \otimes V_{P^i_l}+\sum_{l=D}^{n_{\rm gr}-1} \ket{l}\bra{l} \otimes I, \label{eq:VOF1}
\end{equation}
which acts as
\begin{equation}
    V^{i}_{\rm OF 1} \ket{l}\ket{0}=\ket{l}\ket{P^i_l}
\end{equation}
for any $l\in[D]_0$.

In addition, for any $i\in[d]$ and $l\in[D]_0$, using the controlled $V^l_{\rm set}$ and $V_{P^i_l}$, we construct the following gate on the system of two $m_{\rm gr}$-qubit registers
\begin{equation}
    V^{i,l}_{\rm reset} := (V^l_{\rm set})^\dagger \otimes \ket{P^i_l}\bra{P^i_l} + \sum_{\ket{\psi_\perp}} I \otimes \ket{\psi_\perp}\bra{\psi_\perp}. \label{eq:reset}
\end{equation}
Here, in the second term of the RHS, $\ket{\psi_\perp}$ runs over the $n_{\rm gr}-1$ states that constitute the orthonormal basis of $H$, the Hilbert space on the $m_{\rm gr}$-qubit register, in combination with $\ket{P^i_l}$.
The implementation is as shown in Fig. \ref{fig:reset}.
Note that, in the circuit in this figure, $(V^l_{\rm set})^\dagger$ is activated if the second register takes $\ket{P^i_l}$ but not activated if it takes the state $\ket{\psi}$ orthogonal to $\ket{P^i_l}$, because $(V_{P^i_l})^\dagger\ket{\psi}$ is orthogonal to $(V_{P^i_l})^\dagger\ket{P^i_l}=\ket{0}$.
We then obtain
\begin{equation}
    V^{i}_{\rm OF 2} := \prod_{l=0}^{D-1}V^{i,l}_{\rm reset}=\sum_{l=0}^{D-1} (V^l_{\rm set})^\dagger \otimes \ket{P^i_l}\bra{P^i_l} + \sum_{\ket{\psi_\perp}} I \otimes \ket{\psi_\perp}\bra{\psi_\perp}, \label{eq:VOF2}
\end{equation}
where, in the second sum, $\ket{\psi}_\perp$ runs over the $n_{\rm gr}-D$ states that constitute the orthonormal basis of $H$ in combination with $\ket{P^i_0},...,\ket{P^i_{D-1}}$.

Note that, for any $l\in[D]_0$,
\begin{equation}
    V^{i}_{\rm OF 2}V^{i}_{\rm OF 1}\ket{l}\ket{0}=\ket{0}\ket{P^i_l}
\end{equation}
holds.
Therefore, $V^{i}_{\rm OF 2}V^{i}_{\rm OF 1}$ with a SWAP gate added in the last position is $V^i_{\rm OF}$, with the second register deemed ancillary.

\section{Summary of basic properties of the matrix product state} \label{sec:basic_mps}
The Matrix Product State (MPS) is a representation of a quantum state and, more generally, a high-order tensor, which was first introduced as a way to describe one-dimensional quantum systems efficiently~\cite{Affleck1988,Fannes1992,Ostlund1995,Rommer1997}. It has since become an increasingly popular tool in studying quantum many-body systems. In addition, it has been used to investigate various physical phenomena, including topological phases of matter, quantum phase transitions, and quantum criticality (see review articles~\cite{orus2014practical,okunishi2022developments} and references therein).

As in Sec. \ref{sec:MPS}, let us consider the MPS representation of a tensor $\Psi_{l_1,...,l_d}$ with $d$ indices, each of which runs from $0$ to $D-1$.
Any tensor can always be represented by an MPS, especially the right canonical form of MPS utilized in this paper, by recursively applying the Schmidt/singular-value decomposition as follows.
First, let the degrees of freedom of $\{l_1,\cdots,l_{d-2}\}$ be row components and the degrees of freedom of $\{ l_{d-1},l_{d} \}$ be column components for $\Psi_{l_1,\cdots,l_d}$, and applying the singular-value decomposition (SVD), we obtain
\begin{equation}
\Psi_{l_1,\cdots,l_d} = \sum_{k_{d-2}=1}^{\rho_{d-2}}
V^{1}_{l_1,\cdots,l_{d-2},k_{d-2}} \Lambda^1_{k_{d-2}} U^{d-1}_{k_{d-2},l_{d-1},l_{d}},
\end{equation}
where $\rho_{d-2}:=\min\{D^{d-2},D^2\}$, and left singular vectors $V^{1}_{l_1,\cdots,l_{d-2},k_{d-2}}$, singular values $\Lambda^1_{k_{d-2}}$, and right singular vectors $U^{d-1}_{k_{d-2},l_{d-1},l_{d}}$ for the SVD satisfy following orthonormal conditions 
\begin{align}
\sum_{l_1,\cdots,l_{d-2}=0}^{D-1} \left(V^{1}_{l_1,\cdots,l_{d-2},k_{d-2}}\right)^* V^{1}_{l_1,\cdots,l_{d-2},k^\prime_{d-2}} &= \delta_{k_{d-2},k^\prime_{d-2}}, \\
\sum_{k_{d-2}=1}^{\rho_{d-2}} \left(\Lambda^1_{k_{d-2}}\right)^2 &= \sum_{l_1,\cdots,l_d=0}^{D-1} \left|\Psi_{l_1,...,l_d}\right|^2:=C^2, \\
\sum_{l_{d-1},l_{d}=0}^{D-1} U^{d-1}_{k_{d-2},l_{d-1},l_{d}} \left(U^{d-1}_{k^\prime_{d-2},l_{d-1},l_{d}}\right)^* &= \delta_{k_{d-2},k^\prime_{d-2}}.
\end{align}
Then, we define the coefficients 
\begin{equation}
    \Psi^n_{l_1,\cdots,l_{d-n-1},k_{d-n-1}}=V^{n}_{l_1,\cdots,l_{d-n-1},k_{d-n-1}} \Lambda^n_{k_{d-n-1}}
    \label{eq:psi-n}
\end{equation}
with $n=1$, where $n$ means the number of times SVD is applied.

Second, with $n=2$, letting the degrees of freedom of $\{l_1,\cdots,l_{d-n-1}\}$ be row components and the degrees of freedom of $\{ l_{d-n}, k_{d-n} \}$  be column components for $\Psi^{n-1}$, and applying the singular-value decomposition (SVD) again, we obtain 
\begin{equation}
\Psi^{n-1}_{l_1,\cdots,l_{d-n},k_{d-n}} = \sum_{k_{d-n-1}=1}^{\rho_{d-n-1}}
V^{n}_{l_1,\cdots,l_{d-n-1},k_{d-n-1}} \Lambda^n_{k_{d-n-1}} U^{d-n}_{k_{d-n-1},l_{d-n},k_{d-n}},
\end{equation}
where $\rho_{d-n-1}:=\min\{D^{d-n-1},D^{n+1}\}$, and $V^{n}$, $\Lambda^n$, and $U^{d-n}$ satisfy
\begin{eqnarray}
\sum_{l_1,\cdots,l_{d-n-1}=0}^{D-1} \left(V^{n}_{l_1,\cdots,l_{d-n-1},k_{d-n-1}}\right)^* V^{n}_{l_1,\cdots,l_{d-n-1},k'_{d-n-1}} &=& \delta_{k_{d-n-1},k'_{d-n-1}}, \nonumber \\ \\
\sum_{k_{d-n-1}=1}^{\rho_{d-n-1}} \left(\Lambda^n_{k_{d-n-1}}\right)^2 &=& C^2, \\
\sum_{l_{d-n}=0}^{D-1}\sum_{k_{d-n}=1}^{\rho_{d-n}} U^{d-n}_{k_{d-n-1},l_{d-n},k_{d-n}} \left(U^{d-n}_{k'_{d-n-1},l_{d-n},k_{d-n}}\right)^* &=& \delta_{k_{d-n-1},k'_{d-n-1}}. \nonumber \\ \label{eq:U_for_d-n}
\end{eqnarray}

Repeating the sequence of Eq.~(\ref{eq:psi-n}) to Eq.~(\ref{eq:U_for_d-n}) until $n=d-2$ and defining $U^1_{l_1, k_1} = \Psi^{d-2}_{l_1, k_1}$, which satisfies 
\begin{equation}
\sum_{l_1=0}^{D-1} \sum_{k_1=1}^{\rho_1} |U^1_{l_1, k_1}|^2=C^2,
\end{equation}
we can rewrite $\Psi_{l_1,...,l_d}$ as the right canonical form of MPS 
\begin{equation}
    \Psi_{l_1,...,l_d}=\sum_{k_1=1}^{\rho_1}\cdots\sum_{k_{d-2}=1}^{\rho_{d-2}} U^1_{l_1, k_1} U^2_{k_1,l_2, k_2} \cdots U^{d-1}_{k_{d-2},l_{d-1}, l_{d}},
\end{equation}
and this form is equivalent to Eq.~(\ref{eq:MPS}).
In the practical calculation of MPS, the dimension $\rho_{d-n-1}$ of the matrix $\Lambda^n$, which diverges exponentially with respect to $d$, is truncated to $r$, which we call the bond dimension.
The numerical error $\varepsilon$ that occurs when the truncation of the dimension in the MPS is introduced at the $n$-th SVD during the repeating procedure is the error that appears in the low-rank approximation of the SVD, namely
\begin{equation}
\varepsilon = C^2-\sum_{k_{d-n-1}=1}^{\min\{\rho_{d-n-1},r\}} \left(\Lambda^n_{k_{d-n-1}}\right)^2~.
\end{equation}
Therefore, the shape of the decay function of the singular value $\Lambda^n_{k_{d-n-1}}$, which is a monotonically decreasing non-negative real number for $k_{d-n-1}$, is directly related to the error.
It is known that, in the critical region of one-dimensional quantum systems, this decay function is power-law and $\varepsilon$ thus decays with a power law when we increase $r$ and, accordingly, the DOF in the MPS representation, as seen in Figs.~\ref{fig:dof} and \ref{fig:dof_rand}.

The relationship between MPS and quantum circuits is as described in Sec. \ref{sec:QC}.
We refer to the review articles ~\cite{orus2014practical,okunishi2022developments} and references therein for more detailed properties of MPS itself.

\section{Derivative price as an approximation target function \label{sec:deriv}}

Here, we explain the derivative price, which is considered as an approximation target in the above numerical experiment.

A derivative is a contract between two parties in which the amounts ({\it payoffs}) determined by the prices of some widely traded assets ({\it underlying assets}) such as stocks and bonds are paid and/or received between the parties.
Under some mathematical models that describe the random movement of the underlying asset prices, we can use the established theory to calculate the derivative price (see \cite{hull2003options,shreve2004stochastic}).

In this study, we consider $d$ underlying assets whose prices at time $t$ are denoted by $\vec{S}(t)=(S_1(t),...,S_1(t))$ and obey the Black-Scholes (BS) model \cite{scholes1973pricing,merton1973theory} characterized by the following stochastic differential equation in the risk-neutral measure: for $i\in[d]$,
\begin{equation}
    dS_i(t)=r_{\rm RF}S_i(t)dt+\sigma_i S_i(t) dW_i(t).
\end{equation}
Here, $r_{\rm RF}$ is the real parameter called the risk-free interest rate, and $\sigma_1,...,\sigma_d$ are positive parameters called volatilities.
$W_1,...,W_d$ are Brownian motions and satisfy $dW_idW_j=\rho_{ij}dt$ for $i,j\in[d]$, where the correlation matrix $(\rho_{ij})$ is symmetric and positive definite and satisfies $\rho_{11}=...=\rho_{dd}=1$ and $-1<\rho_{ij}<1$ if $i\ne j$.
Time $t=0$ corresponds to the present.

We consider the derivative in which one party A receives the payoff from the other party B at a predetermined time $T>0$, and its amount $f_{\rm pay}(\vec{S}(T))$ depends on the underlying asset prices at $T$.
Under some technical assumptions, the price of this derivative for A at time $t\in[0,T)$ with $\vec{S}(t)$ being $\vec{s}=(s_1,...,s_d)$ is given by
\begin{equation}
    V(t,\vec{s})=E\left[e^{-r_{\rm RF}(T-t)}f_{\rm pay}(\vec{S}(T)) \ \middle| \ \vec{S}(t)=\vec{s}\right],
\end{equation}
where $E[\cdot]$ denotes the (conditional) expectation in the risk-neutral measure.
This expectation can be calculated by Monte Carlo integration, that is, by generating many sample paths of the time evolution of $\vec{S}(t)$ up to $T$ and averaging $e^{-r_{\rm RF}(T-t)}f_{\rm pay}(\vec{S}(T))$ on the paths.
It is also known that $V(t,\vec{s})$ can be obtained by solving the BS PDE
\begin{eqnarray}
    &&\frac{\partial}{\partial t}V(t,\vec{s}) + \frac{1}{2}\sum_{i,j=1}^d \sigma_i\sigma_j\rho_{ij}s_is_j \frac{\partial^2}{\partial s_i\partial s_j}V(t,\vec{s}) \nonumber \\
    &&\qquad + r_{\rm RF}\left(\sum_{i=1}^d s_i\frac{\partial}{\partial s_i}V(t,\vec{s})-V(t,\vec{s})\right)=0 \label{eq:BSPDE}
\end{eqnarray}
backward from time $T$ to $t$, with the boundary condition in the time direction being
\begin{equation}
    V(T,\vec{s}) = f_{\rm pay}(\vec{s})
\end{equation}
and those in $\vec{s}$ directions set according to the asymptotic behavior of $V(t,\vec{s})$ in the small and large asset price limits.
Solving the BS PDE by quantum computing has been considered in previous studies \cite{miyamoto2021pricing,fontanela2021quantum,gonzalez2021pricing,radha2021quantum,alghassi2022variational}.

Specifically, we consider the {\it worst-of put} option, which has the payoff function 
\begin{equation}
    f_{\rm pay}(\vec{s})=\max\{K-\min\{s_1,...,s_d\},0\},
\end{equation}
and is often incorporated into exotic equity derivatives.
Here, $K$ is a positive constant called the strike.

We then regard $V(0,\vec{s})$ as a function $f(\vec{s})$ and attempt to approximate it.
Because we cannot evaluate this analytically, we compute its values on the grid points using Monte Carlo integration with $10^5$ sample paths and use these values in the numerical experiment.
For this calculation, we used the TF Quant Finance library \cite{tfQuant}.

We set the upper bounds $U_i$ and lower bounds $L_i$ in the $\vec{s}$ space as follows.
$U_i$ is set by
\begin{equation}
    U_i=K\exp\left(\sqrt{2\sigma^2_iT\log\frac{dK}{\epsilon}}\right)
\end{equation}
with $\epsilon=0.01$, following \cite{kangro2000far} on the appropriate grid setting for solving the BS PDE.
On the other hand, we simply set $L_i=0.01K$.

In the numerical experiment described in Sec. \ref{sec:NumExp}, we set $r_{\rm RF}=0,\sigma_1=...=\sigma_5=0.2,K=100,T=1$.
The $10^4$ random sample points in the $\vec{s}$ space used for the numerical experiment were sampled from the distribution of $\vec{S}(t)$ at $t=1$ with the initial value $\vec{S}(0)$ set to $(K,...,K)$.
Therefore, we can regard the worst-of put option under consideration as the option one year after the start, at which it had a 2-year maturity and was at-the-money\footnote{The term ``at-the-money" means the situation where the underlying asset prices are at the marginal point to yield the positive payoff.}.

\bibliography{reference}

\end{document}